\documentstyle[12pt,graphicx]{article}
\begin{document}
\title{Universal temperature corrections to the free energy for the
gravitational field}
\author{G.E. Volovik\\
Low Temperature Laboratory,  Helsinki
     University of Technology\\
    Box 2200, FIN-02015 HUT, Finland\\
and\\
L.D. Landau Institute for Theoretical Physics, \\
117334 Moscow,
     Russia
\\
and\\
A. Zelnikov
\\ Theoretical Physics Institute, University of Alberta
\\ Edmonton, Alberta, Canada
T6G 2J1 \\
and\\
        P.N. Lebedev Physics Institute\\
    Leninskii prospect 53, Moscow 119 991 Russia }

\date{\today}
%modified 11 Sept. 2003
\maketitle
\begin{abstract}
The temperature correction to the free energy of the gravitational field
is considered which does not depend on the Planck energy physics.
The leading correction may be interpreted in terms of the temperature
dependent effective gravitational constant $G_{\rm eff}$.
The temperature correction to $G_{\rm eff}^{-1}$ appears to be
valid for all temperatures $T\ll E_{\rm Planck}$.
It is universal since it is determined only by the number of
fermionic and
bosonic fields with masses $m\ll T$, does not contain the Planck
energy scale $E_{\rm Planck}$ which determines the gravitational constant
at $T=0$, and does not depend on whether or not the gravitational field
obeys the Einstein equations. That is why this universal modification of
the free energy for gravitational field can be used to study
thermodynamics of quantum systems in condensed matter (such as quantum
liquids  superfluid
$^3$He and $^4$He), where the effective gravity emerging for
fermionic and/or bosonic quasiparticles in the low-energy corner is quite
different from the Einstein gravity.
\end{abstract}

\section{Introduction}

The corrections to the Einstein action for the gravitational field
typically depend on the underlying physics of the gravity. There are many
ways  to produce the
gravity: the gravity, of course, can be the fundamental field; but it
also can be an
effective field induced by quantum fluctuations of massive bosonic or fermionic
fields according to the Sakharov scenario
\cite{Sakharov,Jacobson,FrolovFursaev}; it can gradually emerge in the
low-energy corner of
the fermionic system with massless topologically protected fermions
\cite{VolovikBook}; the metric/vierbein fields can emerge also as a
result of the
symmetry breaking as a vacuum expectation value of the bilinear combimations of
spinor fields
\cite{Wetterich}; etc. In most cases the corrections to the Einstein
action are non-universal since they depend on the trans-Planckian physics.
However, there are special cases when the corrections are completely determined
by the infrared physics.   These corrections are universal since they do
not depend on
the details of the underlying physics, and are the same irrespective of whether
the gravity is fundamental or effective.   Here we consider
such universal corrections to the free energy for the gravitational
field which come from the
infrared fermionic and/or bosonic fields whose masses are  small compared
to the temperature $T$.

The same universal terms in the free energy for the effective gravity
arise in the condensed matter systems, in spite of the fact that the main
action  which governs
the effective metric does not resemble the Einstein acton even
remotely. We discuss
such subdominating terms in the bosonic superfluid $^4$He and
fermionic superfluid
$^3$He, where the effective metric field emerges for the bosonic and fermionic
quasiparticles respectively. Since the temperature in these liquids is not very
small compared to the corresponding `Planck' scale $E_{\rm Planck}$ (which is
typically the Debye temperature or the temperature of superfluid
transition $T_c$),
the effect of these terms is accesible for experiments. This allows us to
experimentally verify many issues of the interplay of the gravity and
thermodynamics.

\section{Contribution of massless fermions}
\subsection{Relativistic theory}

Let us start with the modification of the free energy for gravity which
comes from the
massless fermions. Here we are interested in the contribution which
comes from the
thermal relativistic fermions in the background of the curved space.
The free energy of the relativistic gas of fermions in the
stationary gravitational field has the correction containing the
Ricci curvature
${\cal R}$ of the gravitational field. If the gas is in the global
equilibrium with temperature $T_0$ at infinity one obtains in the
high-temperature limit $T^2\gg
\hbar^2{\cal R}$ (but $T\ll E_{\rm Planck}$) the following free energy
\cite{GusevZelnikov,DowkerSchofield2,FrolovZelnikov}:
\begin{equation}
F=F_0 -{7\pi^2N_F\over
360\hbar^3}\int d^3x \sqrt{-g}  T^4 +{N_F\over
288\hbar}\int d^3x \sqrt{-g}  T^2 [{\cal R} + 6 w^2]
~.
\label{FreeEnergyFermionField}
\end{equation}
Here $F_0$ is the temperature independent bulk part of
the free energy of a static gravitational field
\footnote{The main bulk contribution to $F_0$ comes from the space
integral of the Einstein lagrangian taken with the minus sign. 
This follows from the relation of the free energy
to the Euclidean action $F=T I_{\mbox{\tiny Eucl}}$ \cite{York:86}
and the definition of the Euclidean Einstein action
$-{1\over 16 \pi G}\int d^4x \sqrt{g} {\cal
R}$ with the overall minus sign compare to the Einstein action in 
Minkowskian signature
\cite{HawkingHunter}. 
}
\begin{equation}
F_0=-{1\over 16 \pi G}\int d^3x \sqrt{-g} {\cal R}
~.
\label{FreeEnergyGravitynField}
\end{equation}
The $T^4$-term in Eq.(\ref{FreeEnergyFermionField}), $F_4$, is the
thermal free energy
 of massless chiral fermions, where $N_F$ is the number of different
fermionic species (alternatively $N_F=2N_D$ for $N_D$ Dirac fermions with
masses
$m\ll T$). The temperature
$T$ is a local (red-shifted) temperature obeying the Tolman's law
\begin{equation}
T({\bf r})=  {T_0\over \sqrt{|g_{00}({\bf r})|} }~,
\label{TolmanLaw}
\end{equation}
and $T_0={\rm Const}$.

The mixed thermal-gravitational $T^2$-term $F_2$  in
Eq.(\ref{FreeEnergyFermionField}) is the subject of our discussion.
Here $w^2=w^\mu w_\mu~$, where $w_\mu={1\over 2}\partial_\mu\ln
|g_{00}({\bf r})|$ is the 4-acceleration. The temperature dependent part
of the free energy, $F_2+F_4$, is invariant under stationary conformal
transformations
\cite{GusevZelnikov}. This is because the thermodynamics is determined by
the (quasi)particle spectrum, and in the static spacetimes the energy
spectrum $E^2=-p_i p_k g^{ik}/g^{00}$ does not depend on the
conformal factor. The
$6w^2$ term in Eq.(\ref{FreeEnergyFermionField}) provides the
conformal invariance of the $F_2$-term.

The total free energy can be represented in terms of the effective
gravitational `constant' $G_{\rm eff}$
\begin{equation}
F=-{7\pi^2N_F\over
360\hbar^3}\int d^3x \sqrt{-g} \left( T^4 -{15\over 14\pi^2}T^2 w^2\right)
-{1\over 16 \pi }\int d^3x {1\over  G_{\rm eff}}\sqrt{-g} {\cal R} ~,
\label{FreeEnergyFermionField2}
\end{equation}
which is
renormalized due to the
gravitational polarization of matter and becomes coordinate dependent
due to Tolman's
law (\ref{TolmanLaw}):
\begin{equation}
G_{\rm eff}^{-1}({\bf r})-G^{-1}(T=0)=-{\pi N_F\over 18 \hbar}T^2({\bf r})
~.
\label{EffectiveG}
\end{equation}
   Alternatively one could define $G_{\rm
eff}$ as prefactor in front of the curvature terms in the equations
obtained by variation of the free energy $\delta F/\delta g^{\mu\nu}({\bf
r})=0$.
However, this dynamical definition of $G_{\rm eff}$ does not make much
sense for our problems since the effective metric felt by
quasiparticles in condensed matter does not satisfy the Einstein equations
anyway. Thermodynamcal quantities are more robust in this case.
That is why here and
further we determine $G_{\rm eff}^{-1}$ as the prefactor in front of
the curvature ${\cal R}$ in the free energy.

It should be mentioned that Eq.(\ref{FreeEnergyFermionField2}) is valid
for the static (or stationary) case only, since we discuss the system in
a global equilibrium. For the same reason the ${\cal R}T^2$  correction,
in spite of its formally generally-covariant appearance, in fact is not
generally-covariant, since in a global equilibrium the reference frame is
fixed. Just for the same reason the free energy for the gauge field
violates the Lorentz invariance as was discussed for
the static electromagnetic field in Ref. \cite{Fursaev}.

Variation of the free energy over
the stationary metric
$g^{\mu\nu}({\bf r})$ gives the equations for the stationary metric
fields in a global equilibrium, i.e. at fixed
$T_0$. Because of the temperature
corrections to the gravitational `constant',
$G_{\rm eff}$ itself depends on the metric element $g_{00}$.  As a result
the obtained equations do not coincide with the classical Einstein
equations $G_{\mu\nu}=8\pi G_{\rm eff} T^{\rm M}_{\mu\nu}$,  where
$G_{\mu\nu}$ is the Einstein tensor and $T^{\rm M}_{\mu\nu}$ is the
energy--momentum tensor of matter.  Instead one has
\begin{equation}
G_{\mu\nu}=8\pi G \left(T^{\rm M}_{\mu\nu} + T^{{\rm M}+{\rm G}}_{\mu\nu}
\right) ~,
\label{ModifiedEinsteinEquations}
\end{equation}
where $T^{{\rm M}+{\rm G}}_{\mu\nu}$ is non-covariant
mixed thermal-gravitational term which depends both on the temperatute
$T_0$ and on derivatives of the metric tensor.
This means that the gravity in a medium is different from
the gravity in the vacuum: the matter (here the relativistic massless
fermions at given
$T_0$) is not only the source of gravity, but it also modifies the
gravitational interactions which depend on the reference frame in
which the global equilibrium is achieved.

For us it is important that the mixed thermal-gravitational term $F_2$ in
the free energy, and thus the temperature correction to the effective
gravitational constant in Eq.(\ref{EffectiveG}) are the properties of the
low-energy physics of the massless fields in the classical gravitational
background: the $T^2$ correction is universal and does not depend on the
physics at Planck scale. It is determined only by the number and type of
massless fields which are present in the Universe at given temperature,
and these fields  may have nothing to do with the underlying quantum
fields whose vacuum fluctuations  contribute to the
gravitational constant
$G$ at $T=0$ in induced gravity. This means that the temperature
correction does not know whether the gravity is fundamental or effective.

The effect of the polarization of matter in the gravitational field
is extremely
small in normal conditions, since the temperature is extremely small
compared to the
Planck energy scale. The renormalization may become important only when
$g_{00}\rightarrow 0$, i.e. when the   system is close to the
threshold of formation
of the horizon or ergoregion. But in condensed matter the
temperatures are typically
not very low compared with the analogs of 
the Planck energy -- the
Debye temperature
$T_{\rm Debye}\sim 10$ K 
in superfluid $^4$He; the superfluid
transition temperature
$T_c\sim 
1$ mK in $^3$He-A; and the transition temperature
$T_c\sim 10-$100 K 
in superconductors.  That is why this effect can
be pronounced in
the 
effective gravity arising in condensed matter.
Moreover, when the 
thermodynamics or kinetic properties of a condensed matter
system are 
measured, such as the specific heat, entropy or heat
conductivity, 
the
$T=0$ contribution drops out and the temperature dependent terms 
become
dominating. In the high temperature regime 
$T^2\gg
\hbar^2{\cal R}$ the $T^4$ term is dominating in 
thermodynamics. However, the
curvature simulated by textures in 
quantum liquids can be made so
strong that the 
subdominating
$T^2{\cal R}$ term under discussion becomes comparable 
to the $T^4$
term, and in
principle even  the opposite regime 
$T^2\ll
\hbar^2{\cal R}$ can be reached which is called 
the
texture-dominating regime in
condensed matter.

Here we consider 
the analogs of the $T^2{\cal R}$ term on the example of two
condensed 
matter systems.

\subsection{Chiral fermions in $^3$He-A}

Let us start with $^3$He-A   where the
effective metric  felt by chiral fermionic quasiparticles living in the
vicinity of Fermi points (i.e. in the low-energy corner) is \cite{VolovikBook}
\begin{equation}
g_{ij}  ={1\over c_\parallel^2} \hat l^i\hat l^j +{1\over
c_\perp^2}(\delta^{ij} -\hat l^i\hat
l^j)~,~g_{00}=-1~,~\sqrt{-g}={1\over c_\parallel
c_\perp^2}~ .
\label{MetricAPhaseGeneralCovar2}
\end{equation}
Here $\hat{\bf l}$ is the unit vector showing the axis of the uniaxial orbital
anisotropy of the superfluid vacuum;
the longitudinal and transverse `speeds of light' are
\begin{equation}
    c_\parallel={p_F\over m^*}~~,~~ c_\perp={\Delta_0\over p_F}\ll c_\parallel~,
\label{ConnectionLGeneral}
\end{equation}
where $p_F$ and $p_F/m^*$ are Fermi momentum and Fermi velocity
correspondingly, and
$\Delta_0$ is the gap amplitude. We omitted the non-static elements
$g_{0i}$ produced
by the superfluid velocity field, which will be discussed later for the Bose
superfluids, and consider the contribution of only the soft Goldstone
degrees of
freedom related to the deformation of the unit vector $\hat{\bf l}$ (called
textures in condensde matter). That is why the `speeds of light' are
considered as
constants in space, as well as
$g_{00}$, which may be chosen as $g_{00}=-1$. This corresponds to the
ultrastatic spacetime in general relativity.

   Omitting the total derivatives one can calculate
the integral of the Ricci scalar of the metric
in Eq.(\ref{MetricAPhaseGeneralCovar2})
\begin{equation}
\int d^3 x~\sqrt{-g}{\cal R}
=-{1\over 2}\int d^3 x~
c_\parallel c_\perp^2
\left({1\over c_\perp^2}-{1\over c_\parallel^2}\right)^2
((\hat{\bf l}\cdot(
\nabla\times\hat{\bf l}))^2~+\mbox{surface terms}.
\label{CurvaturePart}
\end{equation}

Let us consider how the term like that enters the free energy of $^3$He-A. The
gradient expansion of the free energy in terms of the gradients of $\hat{\bf
l}$-field has been elaborated by Cross in the limit $c_\perp\ll c_\parallel$
\cite{Cross}. At that time it was not known that the effective gravity for
quasiparticles emerges at low energy. But now when we look at the
gradient energy
derived by Cross we find that it does contain the twist term which
quadratically
depends on $T$ \cite{VolovikBook}:
\begin{equation}
   F_2=-{1\over 288\hbar}\int  d^3x { c_\parallel
\over c_\perp^2} T^2  (\hat{\bf l}\cdot(
\nabla\times\hat{\bf l}))^2
\equiv {1\over 144\hbar} \int  d^3x \sqrt{-g}T^2 {\cal R}~.
\label{CorrectionTwistTerm}
\end{equation}
Here $T=T_0$ since $g_{00}=-1$. The thermal part of 
the free energy does
not contain microscopic (trans-Planckian) 
parameters, and is completely
determined by the  Planck 
constant
$\hbar$ and by the parameters of the spectrum of 
fermionic
quasiparticles in the
low-energy limit when the spectrum 
becomes `relativistic':
$E^2=g^{ik}p_ip_k=c_\parallel^2({\bf p}\cdot 
\hat{\bf l})^2
+c_\perp^2({\bf p}\times
\hat{\bf l})^2$ (here  the 
momentum ${\bf p}$ of quasiparticles is
counted from the
Fermi 
point). This term exactly coincides with the
temperature correction 
in Eq.(\ref{FreeEnergyFermionField}) in the
limit 
$c_\perp\ll
c_\parallel$, since the number of the massless chiral 
fermions living in the
low-energy corner of $^3$He-A is
$N_F=2$.
This demonstrates the universality of the $T^2 {\cal R}$-term, i.e.
its independence
of the trans-Planckian physics, while the term $F_0$ is essentially
determined by the trans-Planckian physics \cite{VolovikBook}.

We discussed the contribution to the effective action which comes from massless
fermions. In $^3$He-A there are also analogs of massless photons and
gauge bosons,
moreover in the logarithmic approximation the effective electrodynamics obeys
the same metric $g_{\mu\nu}$ as fermionic quasiparticles, and thus
one could expect
the similar contribution of the gauge bosons to $G_{\rm eff}^{-1}$.
However, this contribution has not been
found. The reason for that is that in the
effective electrodynamics emerging in $^3$He-A the non-renormalizable
non-logarithmic terms are comparable with the covariant terms containing the
logarithmically divergent running coupling \cite{VolovikBook}. As
result the `speed
of light' for photons differs from the  `speed of light' for chiral
fermions. While
$c_\parallel({\rm gauge~bosons})=  c_\parallel({\rm chiral~fermions})$, the
transverse speed of light is essentially bigger:
$c_\perp({\rm gauge~bosons})\gg  c_\perp({\rm chiral~fermions})$. Since
according to Eq.(\ref{CorrectionTwistTerm})
$c_\perp$ is in the denominator, the contribution of the gauge bosons
to the $T^2
{\cal R}$ term is negligible.

The contribution of the scalar fields with their own (acoustic)
metric will be discussed further in the paper.

\section{Contribution of scalar fields}
\subsection{Bosonic field}

The origin of the effective gravity in superfluid $^4$He is
essentially different
from that in $^3$He-A. The superfluid $^3$He-A belongs to the same
universality class
of the fermionic systems as the quantum vacuum of the Standard Model,
and one can
expect that these systems have a common origin of the emerging
gravity represented by
the collective modes related to the deformation of Fermi points. The
superfluid $^4$He is the Bose-liquid where bosonic quasiparticles --
phonons -- play
the role of a scalar field, propagating in the background of the
effective metric
$g_{\mu\nu}$ provided by the moving superfluid condensate (superfluid quantum
vacuum).  The gravity field $g_{\mu\nu}$ experienced by phonons is
simulated mainly
by the superfluid velocity field
${\bf v}_{\rm s}$ (velocity of the superfluid quantum vacuum). The
dynamics of the
velocity field is determined by the hydrodynamic equations instead of
the Einstein
equations. Nevertheless, the  universal
$T^2{\cal R}$ 
term must enter the effective action for $g_{\mu\nu}$,
and this 
term
must be universal, such as in Eq.(\ref{EffectiveG}) but with 
the
prefactor determined
now by the number $N_s$ of scalar fields 
with masses $m\ll 
T$
\cite{GusevZelnikov,DowkerSchofield1}:
\begin{equation}
F=F_0 
-{\pi^2N_s\over
90\hbar^3}\int d^3x \sqrt{-g}  T^4 
-{N_s\over
144\hbar} \int d^3x \sqrt{-g} T^2 [{\cal R}+6 
w^2]
~.
\label{FreeScalarField}
\end{equation}
In general relativity 
this leads to the following modification of
the effective 
gravitational constant determined as the prefactor in front
of the 
curvature in the free energy functional:
\begin{equation}
G_{\rm 
eff}^{-1}-G^{-1}(T=0)={\pi\over 9 
\hbar}N_sT^2
~.
\label{EffectiveGs}
\end{equation}

\subsection{Phonon 
contribution in superfluid $^4$He.}

Phonons obey the following 
effective metric (which is called the
acoustic metric
\cite{Unruh}):
\begin{equation}
\sqrt{-g}={1\over c^3}~~,~~g_{00}=- \left(1-{{\bf v}_{\rm s}^2\over
c^2}\right)~~,~~
g_{ij}= {1\over c^2}\delta_{ij}~~,~~g_{0i}=-{v_{{\rm s}i}\over c^2}~,
\label{MetricHelium4}
\end{equation}
where $c$ is the speed of sound;
${\bf v}_{\rm s}$ is the superfluid velocity; and we neglected the
conformal factor
which depends on the mass density
$\rho$ of the liquid.
Let us consider the simplest case when the
velocity field is stationary. This means that there is a preferred reference
frame, in which the velocity field does not depend on time, and ${\bf v}_{\rm
s}({\bf r})$ is the velocity with respect to this frame. The metric is thus
stationary, but not static. Let us also assume for simplicity that
$c$ and $\rho$
are space independent.  Then the expression for curvature in terms of
the velocity
field is
\cite{FischerVisser}:
\begin{equation}
{\cal R}={1\over 2} (\nabla\times{\bf v}_{\rm s})^2+
\nabla\cdot({\bf v}_{\rm s}(\nabla\cdot{\bf v}_{\rm s}))+
\nabla\cdot(({\bf v}_{\rm s}\cdot\nabla){\bf v}_{\rm s}) .
\label{CurvatureVelocity1}
\end{equation}
For pure rotation ${\bf v}_{\rm s}={\bf \Omega}\times{\bf r}$ the
curvature is zero
${\cal R}=0$, since this corresponds to Minkowski space-time in the
rotating frame.
That is why let us consider the case when the velocity of the
superfluid vacuum is
curl-free ($\nabla\times{\bf v}_{\rm s}=0$), which is the case for
superfluid $^4$He
    when quantized vortices are absent. Then one has
\begin{equation}
{\cal R}=\nabla_i\nabla_k( v_{{\rm s}i} v_{{\rm s}k})~.
\label{CurvatureVelocity2}
\end{equation}
If in addition the superfluid vacuum is incompressible
($\nabla\cdot{\bf
v}_{\rm s}=0$),  the curvature transforms to
\begin{equation}
{\cal R}={1\over 2} \Delta ({\bf v}_{\rm s}^2)={1\over 2} c^2\Delta
g_{00}=-\Delta
\Phi~,
\label{CurvatureHelium4}
\end{equation}
where $\Phi$ is the analog of the gravitational potential:
$g_{00}=-(1+2 \Phi/c^2)$.

Let us apply Eq.(\ref{FreeScalarField}) to the superfluid $^4$He where
$N_s=1$. For simplicity we consider the static effective spacetimes, i.e.
those which satisfy the condition $\nabla\times({\bf
v}_{\rm s}/(c^2-v_{\rm s}^2))=0$. Then integrating by parts one obtains:
\begin{equation}
    F-F_0=\int d^3x \left(-{\pi^2 T^4 \over
90\hbar^3 c^3}   - 
{T^2\over
144 \hbar  c^5}     { \nabla ({\bf
v}_{\rm s}^2)\cdot 
\nabla({\bf
v}_{\rm s}^2) \over 1-{{\bf v}_{\rm s}^2\over  c^2}} 
\right)~.
\label{CorrectionHelium4}
\end{equation}
Let us remind that 
$T({\bf r})=T_0/
\sqrt{1- {\bf v}_{\rm s}^2({\bf r})/ c^2}$ is a 
red-shifted
temperature measured by the local observer who lives in 
the liquid and
uses phonons for communication, while
  $T_0$ is the 
real temperature of the liquid, which is
constant across the 
container in a global equilibrium.

Note that the free energy 
$F_0$ for the effective gravity itself (the
hydrodynamic energy 
of superfluid
$^4$He in terms of
${\bf v}_{\rm s}$  and $\rho$) is 
very different from that in general
relativity, and it certainly 
cannot be represented only in terms of the
curvature term
$\sqrt{-g} 
{\cal R}$. This is because the
hydrodynamic equations being 
determined by the ultra-violet
`Planck-scale' physics
are not obeying 
the effective acoustic metric (\ref{MetricHelium4}).
The same 
actually occurs in $^3$He-A, though the effective gravity there 
is
essentially improved as compared to that in  superfluid
$^4$He, 
and one can even identify some of the components of the 
Einstein
action. However, in both liquids the temperature correction 
to the
curvature term
in Eqs.(\ref{CorrectionHelium4}) and 
(\ref{CorrectionTwistTerm}) is within the
responsibility of the 
emerging infra-red relativistic physics and is
independent of
the 
microscopic (Planck) physics.

\subsection{Fields with different 
effective metrics.}

In the fermionic liquid, the velocity-dependent 
$T^2{\cal R}$ term (such as the
second term in 
Eq.(\ref{CorrectionHelium4})) is also present.
Actually there are 
even
several such terms: one comes from $N_F=2$ chiral
fermions with 
their metric $g_{\mu\nu}^F$ in
Eq.(\ref{MetricAPhaseGeneralCovar2}) 
modified by the superfluid
velocity field (see
Eq.(9.13) in 
\cite{VolovikBook}); another one comes from the massless
scalar field 
of
the sound with its acoustic metric $g_{\mu\nu}^s$ 
in
Eq.(\ref{MetricHelium4}); and
there are also two massless scalar 
fields of spin waves with their own
`spin-acoustic' metric. These 
contributions to the
$T^2$ correction can be combined in the 
following way
\begin{equation}
F_2= {T_0^2\over 288 \hbar}\int 
d^3x\left(\sum_F{\sqrt{-g^F}\over
\vert g_{00}^F\vert 
}\left({\cal
R}_F+6w^2_F\right)-2\sum_s{\sqrt{-g^s}\over 
\vert
g_{00}^s\vert }\left({\cal R}_s+6w^2_s\right)\right) 
~.
\label{EffectiveGfs}
\end{equation}
Here $T_0$ is the real 
temperature in the liquid,
which is constant in a global equilibrium, 
while the local
red-shifted temperature
is different for different 
quasiparticles, since it depends on the
effective metric
experienced 
by a given excitation.

\section{Discussion}

There are many 
condensed matter systems where the low-energy quasiparticles
obey the 
effective Lorentzian metric. The dynamics of this effective 
metric
is, typically, essentially different from the dynamics of 
the
gravitational field
in Einstein theory.  However, the 
thermodynamics appears to be identical, since
it is determined solely 
by the infra-red physics, which does not
depend on whether
the 
gravity is fundamental or effective. This similarity can be useful 
both for
condensed matter and for gravity.

In particular this will 
be useful for the consideration of the low-temperature
limit. The 
procedure of the gradient expansion in quantum liquids
assumes that 
the
gradients are smaller than all other quantities, in particular, 
the
correction to
the action for the metric field in Eq.(\ref{CorrectionTwistTerm}) is
valid only if
the effective curvature is small enough: $\hbar^2{\cal R}\ll T^2\ll E_{\rm
Planck}^2\equiv
\Delta_0^2$, i.e. when the temperture is relatively high.  However, in quantum
liquids or superconductors the opposite (texture-dominating) limit
$T^2\ll\hbar^2{\cal R}\ll  E_{\rm Planck}^2$ is also achievable. Moreover the
thermodynamic effects related to the textures (curvature) become
dominating in this
limit. An example is provided by the  $-T^2\sqrt{{\cal B}}$ term in the free
energy of the quasi-two-dimensional $d$-wave superconductor in applied
magnetic field ${\cal B}$
\cite{DWave}. This term comes from the gapless fermions in the background of
the $U(1)$ field, whose curvature ${\cal B}$ substitutes the Riemann
curvature ${\cal R}$ of the gravitational field. It  determines the
thermodynamics of
    $d$-wave superconductors in the low-$T$ limit
$T^2\ll {\cal B}$, and in particular the specific heat
$C\propto T\sqrt{{\cal B}}$ has been measured in cuprate superconductors
\cite{DWaveExp}.  Whether the corresponding gravitational term
$-T^2\sqrt{\cal R}$
appears  at
$T^2\ll\hbar^2{\cal R}$ in the free energy of quasi-two-dimensional
systems with
effective Lorentzian metric is the problem for future investigations.

Since these effects are determined solely by the infra-red
physics, one can use for their calculations the methods developed in the
relativistic theory. Since these terms are not sensitive to the dynamics of the
effective gravity field, one can assume that this dynamics is governed  by the
Einstein equations. Then the powerful theorems derived for the
Einstein gravity can
be used, including the connection between the curvature terms in the
free energy and
the entropy of the horizon. Thus the condensed matter can serve as
the arena where
the connection between the gravity and thermodynamics can be
exploited and developed.

We are grateful to A.A. Starobinsky for numerous discussions. This work
was supported by ESF COSLAB Programme and by the Russian Foundation for
Fundamental Research. AZ is grateful to the Killam Trust for its
financial support.

\end{document}